\documentclass[a4paper]{jpconf}

\ifpdf%                                % if we use pdflatex
  \pdfoutput=1\relax                   % create PDFs from pdfLaTeX
  \usepackage{graphicx}                % allow us to embed graphics files
  \DeclareGraphicsExtensions{.pdf,.png,.jpg,.jpeg} % for pdflatex we expect .pdf, .png, or .jpg files
\else%                                 % else we use pure latex
  \usepackage{graphicx}                % allow us to embed graphics files
  \DeclareGraphicsExtensions{.eps}     % for pure latex we expect eps files
\fi%

\graphicspath{{figures/}{pictures/}{images/}{./}} % where to search for the images

\usepackage{microtype}                 % use micro-typography (slightly more compact, better to read)
\PassOptionsToPackage{warn}{textcomp}  % to address font issues with \textrightarrow
\usepackage{textcomp}                  % use better special symbols
\usepackage{mathptmx}                  % use matching math font
\usepackage{times}                     % we use Times as the main font
\usepackage{cite}                      % needed to automatically sort the references
\usepackage{tabu}                      % only used for the table example
\usepackage{booktabs}                  % only used for the table example

\usepackage{amssymb}
\newcommand{\mathbold}[1]{\ensuremath{\boldsymbol{\mathbf{#1}}}}

\usepackage{scalefnt,letltxmacro}
\LetLtxMacro{\oldtextsc}{\textsc}
\renewcommand{\textsc}[1]{\oldtextsc{\scalefont{1.10}#1}}
\usepackage[ttdefault=true]{AnonymousPro}

\usepackage{amsmath}

\usepackage
[acronym,smallcaps,nowarn,section,nogroupskip,nonumberlist]{glossaries}
\glsdisablehyper{}

\usepackage{afterpage}
\usepackage{longtable}
\usepackage{lineno}
\usepackage{booktabs, array, tabulary}
\newacronym{VI}{vi}{variational inference}
\newacronym{KL}{kl}{Kullback-Leibler}
\newacronym{ELBO}{elbo}{\emph{evidence lower bound}}
\newacronym{MCMC}{mcmc}{Markov chain Monte Carlo}
\newacronym{GP}{gp}{Gaussian Process}

\newcommand{\mbS}{\mathbold{S}}

\newcommand{\mbY}{\mathbold{Y}}

\begin{document}
\title{Towards forecast techniques for business analysts of large commercial data sets using matrix factorization methods}

\author{Rodrigo Rivera, Ivan Nazarov, Evgeny Burnaev}

\address{Skolkovo Institute of Science and Technology}

\ead{rodrigo.riveracastro@skoltech.ru, ivan.nazarov@skolkovotech.ru, e.burnaev@skoltech.ru}

\begin{abstract}%you must include an abstract
This research article suggests that there are significant benefits in exposing demand planners to forecasting methods using matrix completion techniques. This study aims to contribute to a better understanding of the field of forecasting with multivariate time series prediction by focusing on the dimension of large commercial data sets with hierarchies. This research highlights that there has neither been sufficient academic research in this sub-field nor dissemination among practitioners in the business sector. This study seeks to innovate by presenting a matrix completion method for short-term demand forecast of time series data on relevant commercial problems. Albeit computing intensive, this method outperforms the state of the art while remaining accessible to business users. The object of research is matrix completion for time series in a big data context within the industry. The subject of the research is forecasting product demand using techniques for multivariate hierarchical time series prediction that are both precise and accessible to non-technical business experts. Apart from a methodological innovation, this research seeks to introduce practitioners to novel methods for hierarchical multivariate time series prediction. The research outcome is of interest for organizations requiring precise forecasts yet lacking the appropriate human capital to develop them.
\end{abstract}

\section{Non-technical Summary}\label{section:introduction}
Companies make an effort to collect and evaluate supply chain data with the objective of providing accurate forecasts to support decision-makers in estimating the company performance. To contribute towards this objective, this research presents a method for short-term forecasts based on matrix factorization techniques that is well-suited for demand planners with domain knowledge but limited understanding of machine learning methods. Organizations up-skilling existing personnel into data scientists and seeking to increase forecast accuracy will find value in the proposed methodology.
\section{Research Abstract and Goals}\label{researchandgoals}
\subsection{Introduction}
Commercial sales data is commonly recorded by organizations and serves as the basis for demand forecasting and inventory planning. Nevertheless, retailers still experience out-of-stock (OOS) events with rates amounting to 8.3\% worldwide according to Gruen et al., 2003, \cite{gruen2002retail}. An unintended consequence is that almost half of the customers leave the store without purchasing, while the other half opts for substitution, \cite{gruen2002retail}. Customers finding poor availability in a store on a regular basis will not only incur short-term lost sales but will also decide not to return to the store, \cite{doi:10.1287/mnsc.1060.0577}.
\subsection{Aims and Backgrounds}
The objective of this research is to present an approach to short-term demand forecasting accessible to non-technical business experts offering competitive results comparable to ARIMA. The use of novel machine learning methods for this field is a promising area with little academic research and with insufficient efforts to expose practitioners to them \cite{2017arXiv170905548R}. One example is the work of Carbonneau et al., 2008, \cite{Carbonneau2008} comparing neural networks and support vector machines against naive forecasting, moving average, linear regression, and time series models. Machine learning methods provided better results. Having methods accessible to broader audiences is relevant, over 40\% of forecasters use primarily traditional forecasting methods and many have limited computational skills, \cite{Chase2013}.
\subsection{Significance}
Over the last decades, the complexity and requirements for demand forecasting have increased due to a myriad of factors. Similarly, the incentives to develop robust methodologies are significant. Fleisch and Tellkamp, 2004, \cite{Fleisch2003} observed that for discrepancies as low as 2\%, it is worth investing on improving the accuracy of a forecast. Kaul, 2013, \cite{wipro2013} goes as far as claiming that a 10\% reduction in OOS increases revenue of retailers by up to 0.5\%. 

Yet, companies struggle hiring the adequate personnel to address these tasks. A study conducted by the Asia Pacific Economic Cooperation, 2017, \cite{2017apec}, reported that by 2020, Vietnam is expected to face a shortage of over 500,000 employees with data science and analytics skills. Yet, over 80\% of the 54 million workers in the country do not have the necessary skill set. In Europe, a survey conducted by ESADE Business School, 2018, \cite{2018esade}, reported that over 70\% of surveyed businesses struggled hiring data science personnel and over 60\% are resorting to internal trainings to upgrade the skills of existing business analysts.
\subsection{Research Questions \& Objectives}
The research goal of this study is to propose an approach for time series forecast that can be easily adopted by business practitioners. For this purpose, the study poses the questions: (1) What is the state of the art in academic research of time series prediction with structures? (2) How does the proposed method differs from popular approaches applied to time series prediction tasks? 

To achieve the research goal, two objectives have been assigned: (a) To review the existing theory on multivariate time series prediction and specially on techniques for dynamic hierarchical structures; (b)
To compare the performance of the proposed forecast methods against ARIMAX and gradient boosting. 

The object of research is the balance between accessibility and precision of predictive methods for time series in a big data context within the industry. The subject of the research is forecasting product demand using techniques for multivariate time series prediction with hierarchies.
\section{Literature Review}\label{section:literaturereview}
\subsection{Demand Forecasting}\label{df}
Supply chain management (SCM) in general and demand forecasting in particular are fields that have commanded attention from different communities according to Attar, 2016, \cite{Attar2016}. Thus, a comprehensive treatment can be found in the works of Chase, 2013,\cite{Chase2013} and Gilliand, 2015, \cite{Gilliand2015}.

Sales forecasting is an important part of the supply chain management. A problem that can be stated as predicting future values $x_{t+h}$, given a time series $x_{t-w+1},\ldots, x_{t-1}, x_t$, where $w$ is the length of the window on the time series and $h$ is called the prediction horizon \cite{Manisha2013}. To obtain these predictions, quantitative methods of the statistical family of ARIMA, exponential smoothing models and alike are often used and trained with historical data to forecast future points with the objective of improving the forecasting accuracy. For this purpose, techniques such as decomposition and combination are used. Yet, Ahmed, et al., 2010, \cite{Ahmed2010}, argue that there have been few large scale comparison studies of machine learning models for regression or time series aimed at forecasting problems. In their study, they evaluated multilayer perceptron, Bayesian neural networks, radial basis functions, kernel regression, K-nearest neighbor regression, CART regression trees, support vector regression, and Gaussian processes. The authors identified multilayer perceptron and Gaussian process regression as the best methods. Similarly, researchers from the economics community such as Chernozhukov et al., 2017, \cite{Chernozhukov2017}, have also explored demand prediction through machine learning methods by combining variations of LASSO in the so-called 'double machine learning methods'.

In the retail and manufacturing sectors, there are also cases of demand forecasting such as the work of Tirkeş et al., 2017 \cite{Tirkes2017}. To forecast demand of edibles, they compared Holt-Winters (HW), trend analysis and decomposition models'. These are popular methods in the food and beverage industry together with ARIMA models and time series regression. Using Mean Absolute Percentage Percentage Error (MAPE), the HW and Decomposition models obtained the best results. However, they were not able to identify the conditions that make a method better than the other. Similarly, Taylor, 2007, \cite{Taylor2007}, deals with time series characterized by high volatility skewness and time-varying to forecast daily sales for a supermarket chain at the point of sale. The approach was based on exponential smoothing of the cumulative distribution function, which can be viewed as an extension of generalized exponential smoothing to quantile forecasting. This was justified based on the pervasiveness of exponential smoothing for point forecasting in inventory control. Bianchi, 2017, \cite{Bianchi2017}, experimented with recurrent neural networks for short term forecasting of real-valued time series. While Carbeonneau et al., 2008, \cite{Carbonneau2008}, explored demand forecasting with incomplete information comparing recurrent neural networks and support vector machines against traditional methods such as linear regression or moving average. Their findings suggest that while recurrent neural networks and support vector machines show the best performance, their forecasting accuracy was not statistically significantly better than that of the regression model.

For the electronics manufacturing industry, Wan, 2016, \cite{Wan2016}, introduced SVM regression to the supply chain of various producers. They concluded that it yields good results compared to other forecasting methods. Although SVM regression is a popular method for forecasting, not everyone has identified it as the most effective method. For example, Lu et al., 2012, \cite{Lu2012}, used multivariate adaptive regression splines (MARS), a nonlinear and non-parametric regression methodology, to construct sales forecasting models for computer wholesalers. They argue that artificial neural networks are unable to identify important forecasting variables. Positively, their MARS model outperforms backpropagation neural networks, support vector machines, a cerebellar model articulation controller neural network, an extreme learning machine, an ARIMA model, a multivariate linear regression model, and four two-stage forecasting schemes across various performance criteria. Other manufacturing-centric sectors such as fashion have also delved into demand forecasting but to a different extent. One example is Liu et al., 2013, \cite{Liu2013}. They claim that pure statistical methods are not yet commonplace in the fashion industry. It is preferred to make use of judgmental forecasts or a combination of quantitative and qualitative forecasts.

The forecasting of slow-movers or products with limited historic records is a niche that deserves special attention, as the lack of records affect the performance of the model, specially if there are long periods of zero demand, according to Dolgui, 2005, \cite{Dolgui2005}. Common approaches are the \textit{Croston's method}, where an exponential smoothing function is applied separately to the various nonzero continuous intervals, the use of a Poisson probability distribution as well as of bootstrap methods. Specially challenging is the situation where an item had only zero records in the past, for example in the case of new product introductions. A common solution is to group items hierarchically and obtain an average of the demand distribution.
\subsection{Hierarchical Forecasting}
Forecasting of hierarchical models is a niche in the demand forecasting literature. For Synthetos et al., 2016, \cite{Syntetos2016}, hierarchies enable logical techniques of aggregation and disaggregation such as cross-sectional aggregation and hierarchical product aggregation. In a general hierarchical model, with $K$ levels, level $0$ is defined as the completely aggregated series. Each level from $1$ to $K$ denotes a further disaggregation down to level $K$ containing the most disaggregated, bottom level, time series. Observations are recorded at times $t = 1, \ldots ,n$ and the aim is to make forecasts for each series at each level at times $t = n+1, \ldots ,n+h$.

Let $Y_t$ be the the aggregate of all series at time $t$. In a hierarchical model the observations at higher levels can be obtained by summing the series below. Thus $Y_t = \sum_i Y_{i,t}$, where \textit{i} indexes a generic series at level 1 of the hierarchy, and $Y_{i,t} = \sum_i Y_{ij,t}$ etc.
If we let $\mbY_{i,t}$ denote the vector of all observations at level \textit{i} and time \textit{t} then, 
\begin{equation}
\mbY_t = [Y_t,\mbY_{1,t}, \ldots ,\mbY_{K,t}]'.
\end{equation}
Using the 'summing' matrix, $\mbS$, which stores the structure of the hierarchy, $\mbY_{t}$ can be found from the bottom level series.
\begin{equation}
\mbY_t = \mbS\mbY_{K,t}
\end{equation}
The 'summing' matrix $\mbS$ is a matrix of order $m \times m_K$. Where $m$ is the total number of series in the hierarchy and the number of series in each level is $m_i$ (for $i = 0, 1, \ldots, K$).

Research on hierarchical demand forecasting of sales has been centered on single prediction models for large inventories regardless of hierarchy. Huang et al., 2015, \cite{Huang2015}, advocate thus to experiment with fitting different models to different products arguing that this increases the accuracy. For hierarchical aggregation with intermittent demand, Synthetos et al., 2016, \cite{Syntetos2016}, evaluated with success, if conventional parametric methods such as exponential smoothing or bootstrapping can be used to simulate the entire distribution of demand. 
\subsubsection{Cross-sectional Aggregation}\label{df:hf:cs}
A common technique to handle hierarchical forecasts is through the use of cross-sectional aggregation. It consists of summing demands across items, customers or locations for the same time periods with the objective of extracting useful information from the aggregate. An example of cross-sectional aggregation is to forecast seasonal demand, \cite{Syntetos2016}. In this setting, it is necessary to have a significant amount of observations. However, if the demand is cross-sectionally aggregated, it is possible to determine a seasonal pattern. An alternative is to aggregate products by considering locations, as they might have similar behavioral patterns; for example, stores from the same region. As a minimum of two years of observations is needed to obtain an individual seasonal index (ISI), \cite{Syntetos2016}, a group seasonal index' (GSI) can be used as an alternative. Here, the demand on an item level is aggregated across a group of peers with a common seasonal pattern. Two options are available to generate a GSI: (a) \textit{Withycombe} (WGSI) sums demand across the product group and using the aggregate series proceeds to estimate the seasonal indexes; (b) \textit{Dalhart} (DGSI) calculates individual indexes for each item in the group, averages them across the whole group and uses this average on every item. 

The practitioner can choose from three options: (1) an individual seasonal index for all series or (2) a group approach (GSI) using either a WGSI or a DGSI or (3) decide on an individual basis for each time series. Synthetos et al., \cite{Syntetos2016}, mention that multiple authors sought to identify the advantages of an approach over the other. The results have been inconclusive and depended on the coefficient of variation of the \textit{deseasonalized} series. 
\subsubsection{Hierarchical Product Aggregation}\label{df:hf:hp}
Both Synthetos et al., \cite{Syntetos2016}, and Morgan, 2015, \cite{Morgan2015}, identify two approaches used to deal with hierarchical aggregation forecasting, namely bottom-up and top-down; additionally, a third middle-out approach can also be found in the literature.
\begin{itemize}
\item Bottom-up approach: Forecast at the lowest level of the hierarchy and then aggregate them to obtain an estimate at a higher hierarchical level.
\item Top-down approach: Forecast at an aggregate level and then, disaggregate them to lower hierarchical levels. For example, by multiplying the aggregate demand forecast by a proration factor such as the expected ratio of the respective individual item demand to aggregate demand.
\item Middle-out approach: Forecast on a middle level in the hierarchy and then aggregate bottom-up, as well as disaggregate top-down.
\item Interpersonal reconciliation: Reconcile manually using advice from demand forecasting experts in the organization. 
\end{itemize}
Kamath, 2009, \cite{Kamath2009} mentions that interpersonal reconciliation can be unreliable and lead to inaccurate forecasts. Along these lines, Morgan, 2015, \cite{Morgan2015} discussed the consequences of using combination methods such as optimal combination, bottom-up and top-down. The best method depends largely on the situation at hand, all three methods have been seen to give good results and outperform the others at different times. For example, if the bottom level is noisy using top-down might be preferable. However, if there is strong seasonality in the data, a bottom-up approach could be better suited. Synthetos et al., \cite{Syntetos2016} also support the view that there is no definite approach. Discussions in the literature have been inconclusive. An important determinant of such comparative performance is the hierarchical level (aggregate or sub-aggregate) for which a forecast is needed. Some authors advocate the top down approach arguing that doing bottom-up under-performs as the data is too aggregated to represent diverse demands. A disadvantage of the top-down approach is that commonly historical performance of an item's contribution to the aggregated group is used as a parameter for disaggregation. However, this is subject to sampling variability, it is not possible to estimate the standard deviation at the individual level from the standard deviation at the aggregate level. A top-down approach is specially appropriate whenever there is a change in policy. For example, if the pack sizes are changed. However, in this case, the disaggregation needs to take place through judgmental estimates. 
\subsubsection{Hierarchical Customer or Location Aggregation}\label{df:hf:hc}
Another form of aggregation is through customers or locations. Synthetos et al., 2016, \cite{Syntetos2016}, argue that here also there has not been a method that can be identified as superior. In the literature is possible to find approaches using sophisticated clustering methodologies but they did not fare better against random grouping. However, the appropriate level of aggregation has been shown to be determinant in the performance of the forecast accuracy. There are both arguments in favor of a granular aggregation as well as of a broader one. For example, significant aggregation leads to lower sampling error but risks mixing seasonality patterns from two very different products.

Finally, one approach going beyond the 'traditional' grouping methods for hierarchical forecasting of time series is to lever Granger causality. For example, Lemmens et al., 2008, \cite{Lemmens2008}, focused on decomposing Granger causality over the spectrum.

\section{Matrix Factorization Method}\label{section:mfm}
Matrix factorization methods (MF) are used in a variety of applications such as recommender systems, signal processing~\cite{Weng2012}, computer vision~\cite{Chen2004} and others.

Let $Y$ be $T\times n$ matrix of observations of $n$ objects spanning the period of $T$ time steps, i.e. each column $i=1,\,\ldots,\,n$ of $Y$ is a times series $y^{(i)} = (Y_{ti})_{t=1}^T$ related to the $i$-th object. For instance, $Y$ may represent consumption expenditures within a longitudinal study of households, the hourly records of electricity consumption at different substations, financial time series or changes in stock levels.

The problem of factorizing a fully or partially observed $T\times n$ matrix $Y$ consists of finding $d$-dimensional factors $X$ and the corresponding factor loadings $F$, in the form of $T \times d$ and $d \times n$ matrices respectively, such that their product $X F$ most accurately recovers the observed $Y$, i.e. $Y_{ti} \approx \sum_{j=1}^d X_{tj} F_{ji}$. This is usually achieved by solving the following optimization problem:
\begin{equation} \label{eq:general_mf}
\begin{aligned}
& \underset{F, X}{\text{minimize}}
  & & \tfrac1{2 T n} \|Y - X F\|^2_F + \lambda_F \mathcal{R}_F(F) + \lambda_X \mathcal{R}_X(X)
      \,,
\end{aligned}
\end{equation}
where $\lambda_F$ and $\lambda_X$ are non-negative regularization coefficients which govern the trade-off between the reconstruction error and the regularizing terms $\mathcal{R}_F$ and $\mathcal{R}_X$. The latter depend on the particular desired properties of the factorization, such as sparsity or row-wise group
sparsity~\cite{nazarovetal2018}, typically in conjunction with a Ridge regression-type penalty ($\ell^2$ norm).

The key issue with~\eqref{eq:general_mf} is that without extra structural requirements on $X$ it is impossible to apply this technique to time series prediction. A recent paper~\cite{yuetal2016} proposes a novel regularization term $\mathcal{R}_X$ for~\eqref{eq:general_mf}, that enables forecasting beyond $T$ by imposing autoregressive time-series properties on the latent factors $X$. The corresponding optimization problem which imposes $AR(p)$ (autoregression of order $p$) dynamics on the factors is
\begin{equation} \label{eq:trmf}
\begin{aligned}
& \underset{F, X, \theta}{\text{minimize}}
  & & \tfrac1{2 T n} \|Y - X F\|^2_F
      + \tfrac{\lambda_F}2 \tfrac1{d n} \|F\|^2_F
      + \tfrac{\lambda_\theta}2 \tfrac1{d p} \|\theta \|^2_F
      \\
& & & + \tfrac{\lambda_X}2 \Bigl(
            (1 - \eta_X) \tfrac1{T d} \|X\|^2_F
            + \eta_X \tfrac1{(T - p) d}
            \sum_{j=1}^d \sum_{t=p+1}^T \bigl(
                X_{tj} - \sum_{i=1}^p \theta_{ji} X_{t-i,j}
            \bigr)^2
        \Bigr)
      \,.
\end{aligned}
\end{equation}
In this formulation the $\lambda_\theta$ regularizes and stabilizes the estimates of autoregression coefficients $\theta$, that define the dynamics of the latent factor $X$, which are $d$-dimensional
time series. The parameter $\eta_X\in [0, 1]$ regulates the relative contribution of the ridge-like and forecastability penalties to the estimation of the factor series $X$.

The problem~\eqref{eq:trmf} or, in general, any biconvex problem of the form~\eqref{eq:general_mf}, is solved numerically using coordinate descent algorithms that alternates between minimizing the objective with respect to each matrix until some convergence criteria are met.

Forecasting beyond the last observation in $Y$ is done using the estimated latent factors $X$ and the parameters $\theta$ of their autoregressive dynamics. If $x^{(j)} = (X_{tj})_{t=1}^T$ is the time series of the $j$-th latent factor, then its $h$-step ahead dynamic forecast beyond time $T$ is calculated using
\begin{equation} \label{eq:factor_dyn_forecast}
\hat{x}^{(j)}_{T+h\mid T}
	= \theta_{j1} \hat{x}^{(j)}_{T+h-1\mid T}
    + \cdots + \theta_{jp} \hat{x}^{(j)}_{T+h-p\mid T}
	\,,
\end{equation}
with $\hat{x}^{(j)}_{T+h-k\mid T} = x^{(j)}_{T+h-k}$ for any $k \geq h$. Based on~\eqref{eq:factor_dyn_forecast} and the form of the matrix factorization, the $h$-step ahead forecast for the $i$-th object's time series is given by
\begin{equation} \label{eq:value_forecast}
	\hat{y}^{(i)}_{T+h\mid T}
    	= \sum_{j=1}^d \hat{x}^{(j)}_{T+h\mid T} F_{ji}
	\,,
\end{equation}
It is worth noting, that according to~\cite{yuetal2016} the problem~\eqref{eq:trmf} could be further modified to incorporate hierarchical relations among the columns of $Y$ by introducing a so called ``graph regularization'' on the loadings $F$. If a directed graph $G$ represents the {\it ``being necessary for''} binary relation on objects associated with each time series in $y^{(i)}$, then the regularization term on the loadings $F$ becomes
\begin{equation} \label{eq:trmf_graph_loadings}
\tfrac{\lambda_F}2 \Bigl(
	(1 - \eta_F) \tfrac1{d n} \|F\|^2_F
	+ \eta_F \tfrac1{d n} \sum_{i=1}^n
    	\Bigl\| F_{\cdot i}
        	- \tfrac1{\#\{ j\colon (i,j) \in G \}} \sum_{j\colon (i,j) \in G} F_{\cdot j} W_{ij} \Bigr\|^2
\Bigr) \,,
\end{equation}
where $F_{\cdot i}$ is the $i$-th column of the $d\times n$ matrix $F$ of factor loadings and $W_{ij}$ is a positive weight of the link $i \to j$. The parameter $\eta_F \in [0, 1]$ tunes the ridge-like penalty and hierarchical regularization.

\section{Experiments}\label{section:experiments}
To validate the proposed approach, experiments on two different data sets were conducted. A proprietary data set corresponding to an international manufacturer and a subset of the public data set of the Grupo Bimbo Inventory Demand Competition \footnote{https://www.kaggle.com/c/grupo-bimbo-inventory-demand}. An overview of the data sets can be found in table \ref{tab:1}
\begin{table}
 \caption{Overview of data sets used for experiments}
\begin{center}\label{tab:1}
\begin{tabulary}{\linewidth}{CCCC}
    \toprule
    Data set & Type & Periods & Hierarchy  \\ \hline
   Proprietary & 850 Products & Up to 45 periods with 4 delivery dates & Dynamic, up to 7  \\
   Bimbo & 5000 POS & 7 weeks & None \\ \bottomrule
 \end{tabulary}
 \end{center}
 \end{table}
Both data sets reflect common cases in demand forecasting. The Bimbo data set contains 5000 points of sale (POS), each with 7 observations, and the hierarchical dependencies among them is tenuous. Thus, extracting seasonality or leveraging data from other points of sale is difficult. On the other side, the proprietary data can have hierarchies of varying depth up to 7 nodes for each of the 850 products and the parent's demand can influence the child's and vice-versa. The number of periods is higher than in the Bimbo data set. However, many products have sparse demands of only one or two observations and the distributions between the train and the test set can vary.

To compare the proposed method, two popular techniques for demand forecasting were chosen: ARIMAX and an ensemble of gradient boosting. The first is commonly used in the industry by business practitioners, whereas the second has found popularity in machine learning competitions such as Kaggle \footnote{http://www.kaggle.com}. For each POS or product, a new model is trained and tested. Further, in ARIMAX only the time periods and the observed quantities are used to predict. In the case of the proprietary data set the AR parameter was set to 2 and the MA to 3. For the Bimbo data, both AR and MA were set to 2. The quantities were not scaled. Meanwhile, the ensemble consists of four instances of gradient boosting and a final linear regression. For both data sets, features were engineered and scaled using z-score. The proprietary data set contained over 360 features. 
The objective was to mimic the expected actions of two experts with different profiles. On one side, the business expert would choose ARIMAX, whereas a practitioner immersed in Kaggle competitions would carry out extensive feature engineering and apply a complex model.

The Symmetric Mean Absolute Percent Error (SMAPE) is used to evaluate the performance of the models and is defined as following:
\begin{equation}
\text{SMAPE} = \frac{100\%}{n} \sum_{t=1}^n \frac{|F_t-A_t|}{|A_t|+|F_t|}
\end{equation}
To normalize the results, the median of the min-max scaler of SMAPE is used and aggregated over the weeks (Bimbo) or the delivery dates (proprietary data set). The results in table \ref{tab:3} are rounded to the second decimal and a detailed overview for the results of the matrix completion method is presented in table \ref{tab:4}.
\begin{table}
\caption{Overview of results using min-max median SMAPE}
\begin{center}\label{tab:3}
\begin{tabulary}{\linewidth}{CCCCC}
    \toprule
    Data set & Period & ARIMAX & Ensemble & Matrix Completion  \\ \hline
   Bimbo & Week 5 & \textbf{0.20} & 0.28 & NA \\ 
   Bimbo & Week 6 & 0.57 & \textbf{0.50}  & NA  \\ 
   Bimbo & Week 7 & 1.0 & 0.57 & \textbf{0.28}  \\ 
   Proprietary & Delivery date 0 & 0.55  & 0.50 & \textbf{0.32} \\
   Proprietary & Delivery date 1 & 0.54  & 0.51 & \textbf{0.35} \\ 
   Proprietary & Delivery date 2 & 0.65  & \textbf{0.56} & 0.81 \\ 
   Proprietary & Delivery date 3 & 0.64  & \textbf{0.59} & NA \\ \bottomrule
 \end{tabulary}
 \end{center}
 \end{table}
\begin{table}
\caption{Best models for matrix completion according to number of factors and AR-order, see~\eqref{eq:trmf}}
\begin{center}\label{tab:4}
\begin{tabulary}{\linewidth}{lrrrr}
  \toprule
  Nr. Factors ($d$) &       10  &       25  &       50  &       100 \\ 
  Nr. Order ($p$)   &           &           &           &           \\
  \midrule
       $1$          &  0.309371 &  0.309883 &  0.374708 &  0.363766 \\ 
       $3$          &  0.301795 &  0.305454 &  0.310105 &  0.313307 \\ 
       $6$          &  0.311014 &  0.315699 &  0.321317 &  0.321263 \\
       $12$         &  0.335601 &  0.331765 &  0.358228 &  0.362700 \\
  \bottomrule
\end{tabulary}
 \end{center}
 \end{table}
\section{Discussion}
 It is possible to see from the results that each method has its own strengths and uses. ARIMAX is straightforward to use, as it has few parameters and it can be applied directly to the data. Even without any preprocessing, the results can be comparable to advanced methods, specially one step ahead, $x_{t+1}$. As a result, business analysts with little knowledge of machine learning methods or computing can implement it. Similarly, data scientists can use it as a baseline without having a thorough understanding of the business problem. On the other hand, the ensemble shows consistent results even for values far ahead in the future such as $x_{t+3}$. Yet, to achieve good results, a significant effort is required in developing meaningful features. Thus, the ensemble requires a good understanding of machine learning methods as well as of the problem itself. This limits significantly the users within a company that can implement it or forces companies to have two individuals assigned to the problem, a business expert and a data scientist.
On the other hand, the proposed method using matrix factorization excelled for the first periods and then its performance dropped. It obtained better results than both ARIMAX and the ensemble. However, this came at the expense of significant computational time. It required several days to generate predictions for the Bimbo data set. In addition, it does not perform well with short training periods. Thus, it was only possible to generate predictions for the last week. Further, on the proprietary data set, the delivery date 2 has very sparse data. As a consequence, the performance of the method suffered and fared worse than both ARIMAX and the ensemble. Yet, the matrix factorization method achieved its objective of striking a good balance between accurate predictions and accessibility to the business practitioner.
\section{Conclusion}
 This research presented the argument of the relevancy within the machine learning community to develop predictive methods accessible to broader audiences with limited understanding of predictive methods. Companies need accurate forecasts. Yet, they struggle acquiring the appropriate human capital. Existing methods suffer either from being simplistic and thus obtaining sub-optimal results or from being unfit to address modern problems in demand prediction such as time series with hierarchical dependencies and sparse data or from being inaccessible to a business user. This research presented a method that can be easily adopted by business analysts and can deliver state of the art results specially for short-term forecasting. One of its downsides is the computational time. Extending the matrix completion method to very large data sets, to long-term forecasting and to optimize its computing time are some of the most promising research directions for this methodology. 

\ack%this is an unnumbered acknowledgement section
The research in sections \ref{section:introduction}, \ref{section:literaturereview} and \ref{section:mfm} was supported solely by the Ministry of Education and Science of Russian Federation, grant No.14.606.21.0004, grant code: RFMEFI60617X0004. The research presented in other sections was supported by the Mexican National Council for Science and Technology (CONACYT), 2018-000009-01EXTF-00154. Thanks to Evgenii Egorov for the intellectual exchange and suggestions.

\bibliographystyle{vancouver}

\section{Bibliography}
\bibliography{bibliography}

\begin{thebibliography}{10}

\bibitem{gruen2002retail}
Gruen TW, Corsten DS, Bharadwaj S.
\newblock Retail Out of Stocks: A Worldwide Examination of Extent, Causes, and
  Consumer Responses. 2002;.

\bibitem{doi:10.1287/mnsc.1060.0577}
Anderson ET, Fitzsimons GJ, Simester D.
\newblock Measuring and Mitigating the Costs of Stockouts.
\newblock Management Science. 2006;52(11):1751--1763.

\bibitem{2017arXiv170905548R}
{Rivera} R, {Burnaev} E.
\newblock {Forecasting of commercial sales with large scale Gaussian
  Processes}.
\newblock ArXiv e-prints. 2017 Sep;.

\bibitem{Carbonneau2008}
Carbonneau R, Laframboise K, Vahidov R.
\newblock {Application of machine learning techniques for supply chain demand
  forecasting}.
\newblock European Journal of Operational Research. 2008 feb;184(3):1140--1154.

\bibitem{Chase2013}
Chase CW.
\newblock {Demand-Driven Forecasting}.
\newblock Hoboken, NJ, USA: John Wiley {\&} Sons, Inc.; 2013.

\bibitem{Fleisch2003}
Fleisch E, Tellkamp C.
\newblock {Inventory inaccuracy and supply chain performance: a simulation
  study of a retail supply chain}.
\newblock International Journal of Production Economics. 2005
  mar;95(3):373--385.

\bibitem{wipro2013}
Kaul R.
\newblock Retail Out-of-Stock Management: An Outcome-Based Approach. 2013;.

\bibitem{2017apec}
{Pompa} C, {Burke} T.
\newblock {Data Science and Analytics Skills Shortage: Equipping the APEC
  Workforce with the Competencies Demanded by Employers}.
\newblock APEC Human Resource Development Working Group. 2017;.

\bibitem{2018esade}
{Agell} N, {Carricano} M.
\newblock {Adopcion e impacto del Big Data y Advanced Analytics en España}.
\newblock ESADE Business and Law School. 2018 May;.

\bibitem{Attar2016}
Attar KS.
\newblock {Regression for Demand Forecasting}. 2016;5(1).

\bibitem{Gilliand2015}
Gilliand M.
\newblock {Business Forecasting}.
\newblock Gilliland M, Tashman L, Sglavo U, editors. Hoboken, NJ, USA: John
  Wiley {\&} Sons, Inc.; 2015.

\bibitem{Manisha2013}
Manisha G, Vijayalakshmi M.
\newblock {Inter Time Series Sales Forecasting}.
\newblock Jascse. 2013;2(1):55--66.

\bibitem{Ahmed2010}
Ahmed NK, Atiya AF, {El Gayar} N, El-Shishiny H.
\newblock {An empirical comparison of machine learning models for time series
  forecasting}.
\newblock Econometric Reviews. 2010;29(5):594--621.

\bibitem{Chernozhukov2017}
Chernozhukov V, Goldman M, Semenova V, Taddy M.
\newblock {Orthogonal Machine Learning for Demand Estimation: High Dimensional
  Causal Inference in Dynamic Panels}. 2017;.

\bibitem{Tirkes2017}
Tirkeş G, G{\"{u}}ray C, {\c{C}}elebi N.
\newblock {Demand forecasting: a comparison between the Holt-Winters, trend
  analysis and decomposition models}.
\newblock Tehnicki vjesnik - Technical Gazette. 2017 sep;24(Supplement
  2):503--509.

\bibitem{Taylor2007}
Taylor JW.
\newblock {Forecasting daily supermarket sales using exponentially weighted
  quantile regression}.
\newblock European Journal of Operational Research. 2007 apr;178(1):154--167.

\bibitem{Bianchi2017}
Bianchi FM, Maiorino E, Kampffmeyer MC, Rizzi A, Jenssen R.
\newblock {An overview and comparative analysis of Recurrent Neural Networks
  for Short Term Load Forecasting}. 2017 may;.

\bibitem{Wan2016}
Wan Xl, Zhang Z, Rong Xx, Meng Qc.
\newblock {Exploring an Interactive Value-Adding Data-Driven Model of Consumer
  Electronics Supply Chain Based on Least Squares Support Vector Machine}.
\newblock Scientific Programming. 2016;2016:1--13.

\bibitem{Lu2012}
Lu CJ, Lee TS, Lian CM.
\newblock {Sales forecasting for computer wholesalers: A comparison of
  multivariate adaptive regression splines and artificial neural networks}.
\newblock Decision Support Systems. 2012 dec;54(1):584--596.

\bibitem{Liu2013}
Liu N, Ren S, Choi TM, Hui CL, Ng SF.
\newblock {Sales Forecasting for Fashion Retailing Service Industry: A Review}.
\newblock Mathematical Problems in Engineering. 2013;2013:1--9.

\bibitem{Dolgui2005}
Dolgui A, Pashkevich M.
\newblock {Extended beta-binomial model for demand forecasting of multiple
  slow-moving items with low consumption and short requests history}; 2005.

\bibitem{Syntetos2016}
Syntetos AA, Babai Z, Boylan JE, Kolassa S, Nikolopoulos K.
\newblock {Supply chain forecasting: Theory, practice, their gap and the
  future}.
\newblock European Journal of Operational Research. 2016;252(1):1--26.

\bibitem{Huang2015}
Huang W, Zhang Q, Xu W, Fu H, Wang M, Liang X.
\newblock {A Novel Trigger Model for Sales Prediction with Data Mining
  Techniques}.
\newblock Data Science Journal. 2015 may;14(15):15.

\bibitem{Morgan2015}
Morgan L.
\newblock {Forecasting in Hierarchical Models}. 2015;.

\bibitem{Kamath2009}
Kamath S, Jadhwani AC.
\newblock {Demand Forecasting in Apparel Industry in UAE}.
\newblock SSRN Electronic Journal. 2009;.

\bibitem{Lemmens2008}
Lemmens A, Croux C, Dekimpe MG.
\newblock {Measuring and testing Granger causality over the spectrum: An
  application to European production expectation surveys}.
\newblock International Journal of Forecasting. 2008;24(3):414--431.

\bibitem{Weng2012}
Weng Z, Wang X.
\newblock Low-rank matrix completion for array signal processing.
\newblock In: Acoustics, Speech and Signal Processing (ICASSP), 2012 IEEE
  International Conference on. IEEE; 2012. p. 2697--2700.

\bibitem{Chen2004}
Chen P, Suter D.
\newblock Recovering the missing components in a large noisy low-rank matrix:
  Application to SFM.
\newblock IEEE transactions on pattern analysis and machine intelligence.
  2004;26(8):1051--1063.

\bibitem{nazarovetal2018}
{Nazarov} I, {Shirokikh} B, {Burkina} M, {Fedonin} G, {Panov} M.
\newblock {Sparse Group Inductive Matrix Completion}.
\newblock ArXiv e-prints. 2018 Apr;.

\bibitem{yuetal2016}
Yu HF, Rao N, Dhillon IS.
\newblock Temporal Regularized Matrix Factorization for High-dimensional Time
  Series Prediction.
\newblock In: Lee DD, Sugiyama M, Luxburg UV, Guyon I, Garnett R, editors.
  Advances in Neural Information Processing Systems 29. Curran Associates,
  Inc.; 2016. p. 847--855.

\end{thebibliography}

\end{document}